\documentclass[aps,prb,twocolumn,superscriptaddress,groupedaddress]{revtex4} 

\usepackage[dvipdfmx]{graphicx} 
\usepackage{dcolumn} 
\usepackage{amsmath,txfonts,bm}

\usepackage{color} 

\newcommand{\bra}  {\langle}
\newcommand{\ket}  {\rangle}

\begin{document}

\widetext

\title{ Superconductivity arising from layer-differentiation in multi-layer cuprates } 

\author{Kazutaka Nishiguchi} 
\affiliation{Graduate School of Engineering Science, Osaka University, Toyonaka, Osaka 560-8531, Japan} 

\author{Shingo Teranishi} 
\affiliation{Graduate School of Engineering Science, Osaka University, Toyonaka, Osaka 560-8531, Japan} 

\author{Koichi Kusakabe} 
\affiliation{Graduate School of Engineering Science, Osaka University, Toyonaka, Osaka 560-8531, Japan} 

\author{Hideo Aoki} 
\affiliation{Department of Physics, The University of Tokyo, Hongo, Tokyo 113-0033, Japan} 
\affiliation{National Institute of Advanced Industrial Science and Technology (AIST), Tsukuba 305-8568, Japan}

\date{\today}

\begin{abstract} 
In order to theoretically identify the factors governing superconductivity in multi-layer cuprates, 
a three-layer Hubbard model is studied with the two-particle self-consistent (TPSC) approach so as to incorporate electron correlations. 
The linearized Eliashberg equation is then solved for the gap function in a matrix form 
to resolve the role of outer CuO$_2$ planes (OPs) and inner plane (IP). 
We show that OPs dominate IP in the $d_{x^{2}-y^{2}}$-wave superconductivity, 
while IP dominates in the antiferromagnetism. 
This comes from an electron correlation effect in that 
the correlation makes the doping rates different between OPs and IP (i.e., a self-doping effect), 
which occurs in intermediate and strong correlation regimes. 
Namely, the antiferromagnetic fluctuations in IP are stronger due to a stronger electron correlation, 
which simultaneously reduces the quasiparticle density of states in IP with a suppressed $d_{x^{2}-y^{2}}$-wave superconductivity. 
Intriguingly, 
while the off-diagonal (inter-layer) elements in the gap function matrix are tiny, 
{\it inter-layer pair scattering} processes are in fact at work 
in enhancing the superconducting transition temperature $T_{\text{c}}$ through the inter-layer Green's functions.  
This actually causes the trilayer system to have higher $T_{\text{c}}$ than the single-layer in a weak- and intermediate-coupling regimes. 
This picture holds for a range of the on-site Hubbard repulsion $U$ that contains those estimated for the cuprates. 
The present result is qualitatively consistent with nuclear magnetic resonance experiments in multi-layer cuprates superconductors. 
\end{abstract}

\maketitle

\section{Introduction} 
Despite a long history exceeding three decades, 
the high-$T_{\text{c}}$ superconductivity, one of the central interests in condensed matter physics~\cite{Scalapino12,Aoki12}, 
still harbors a host of important questions.  
One salient feature in cuprate superconductors is that, 
if we look at representative homologous series, e.g., Hg-based multi-layer cuprates HgBa$_2$Ca$_{n-1}$Cu$_n$O$_{2n+2+\delta}$ [called Hg-12$(n-1)n$], 
where $n$ is the number of CuO$_2$ layers within a unit cell and $\delta$ is the doping, 
the superconducting (SC) transition temperature $T_{\text{c}}$ becomes the highest for multi-layer cases, 
which possess the highest $T_{\text{c}}$ ($\simeq 135$ K for Hg-1223) to date at ambient pressure~\cite{Schilling93,Takeshita13,Yamamoto15}. 
The CuO$_2$ plane can be described by the Hubbard model with on-site Coulomb repulsion along with electron hopping, 
where a competition between the itinerancy and localization of electrons takes place due to electron correlations. 

If we look more closely at the $n$-layer cuprates, 
$T_{\text{c}}$ systematically depends on $n$ for each homologous series:~\cite{Leggett06TB,Mukuda12} 
$T_{\text{c}}$ increases for $1 \leq n \leq 3$ and decreases slightly and saturates for $n \geq 3$. 
To explain the superconductivity and other electronic properties, 
several pictures for the multi-layer superconductor have been theoretically proposed so far, 
among which are 
an inter-layer Josephson coupling arising from second-order processes of the inter-layer single-electron hopping,~\cite{Anderson97TB,Chakravarty93} 
an inter-layer Josephson pair-tunneling in a macroscopic Ginzburg--Landau scheme,~\cite{Chakravarty04} 
a Coulomb energy saving in the $c$-axis structure,~\cite{Leggett98,Leggett99} 
superconductivity enhanced in artificial superlattices comprising underdoped and overdoped layers,~\cite{Okamoto08,Chen10} 
and an inter-layer pair-hopping arising from higher-order processes of the Coulomb interaction.~\cite{Kusakabe09,Kusakabe12,KN13} 

On an experimental side, 
nuclear magnetic resonance (NMR) experiments exhibit layer-resolved results, 
where we can distinguish between the outer CuO$_2$ planes (OPs) and inner planes (IPs) in multi-layer cuprates. 
Thus the trilayer system is not only the case of highest $T_{\text{c}}$ but also the simplest case accommodating OP and IP. 
The NMR experiments~\cite{Mukuda12,Mukuda08,Mukuda06,Kotegawa01} have in particular shown that 
the carrier concentration is different between OP and IP with more hole (electron) carriers in OP (IP), 
which causes different electronic properties between OP and IP: 
the antiferromagnetic (AF) moments in IPs are much larger than those in OPs, 
and the antiferromagnetism coexists with the superconductivity in the IPs even in the optimally doped regions. 
It is further observed by resolving the OP and IP components that 
the SC gaps seem to develop in two steps where the bulk $T_{\text{c}}$ is determined by a higher $T_{\text{c}}$ in OP 
while IP has a proximity effect from OP up to the overdoped regime. 
Thus the OP seems to dominate the $d_{x^{2}-y^{2}}$ superconductivity, 
while IP the antiferromagnetism in multi-layer cuprates. 
These different behaviors between OP and IP have yet to be theoretically understood, 
and we are still in need of a microscopic theory. 

Recently, 
we have studied normal properties (carrier concentrations and magnetism) of OP and IP 
by investigating the three-layer Hubbard model as an effective model for Hg-1223, 
where we have employed the two-particle self-consistent (TPSC) approach for multi-layer systems.~\cite{KN17} 
The TPSC approach, originally proposed by Vilk and Tremblay,~\cite{Vilk94,Vilk97,Tremblay11} 
is a weak- and intermediate-coupling theory 
in which the spin and charge susceptibilities, along with the double occupancy, are determined self-consistently by assuming the TPSC ansatz, 
and then the self-energy and Green's function can be evaluated. 
When we applied this to the trilayer system, 
we first found that 
the concentration of 
hole carriers tends to be larger in OP than in IP with increasing on-site Coulomb repulsion, 
which is an electron correlation effect. 
Then the AF instability in the IP is shown to be always larger than in the OP. 
These results are consistent with the NMR results on the antiferromagnetism and carrier concentrations in OP and IP mentioned above. 
In particular, 
the many-body charge transfer between the OP and IP can be called a {\it self-doping effect}. 

These have motivated us here to investigate the superconductivity itself in the multi-layer cuprates. 
Thus the present paper theoretically identifies the factors governing superconductivity in multi-layer cuprates 
with a three-layer Hubbard model in the TPSC approach. 
By solving the linearized Eliashberg equation for the gap function in a matrix form to examine the role of OP and IP, 
we shall show for the trilayer strongly correlated system that 
OPs dominate in the $d_{x^{2}-y^{2}}$-wave superconductivity, 
while IP dominates in the antiferromagnetism. 
This is caused by electron correlations 
because the crucial factor for the differentiated doping rates between OPs and IP, i.e. the self-doping effect, 
takes place in intermediate and strong correlation regimes ($U \gtrsim 2$ eV). 
 
Physically, 
the self-doping makes the strengths of electron correlation different between OP and IP: 
the AF spin fluctuations in the IP are stronger than in the OP due to the layer filling closer to half-filling in the IP, 
while the quasiparticle density of states (DOS) is reduced for the same reason around the antinodal regions in the IP, 
suppressing the $d_{x^{2}-y^{2}}$-wave superconductivity. 
We also show that, 
although the off-diagonal (inter-layer) elements in the gap function matrix are tiny, 
the inter-layer pair scattering processes are actually at work in enhancing $T_{\text{c}}$ 
by comparing the results when these processes are turned on and off. 
We further reveal that 
the trilayer system has higher $T_{\text{c}}$ than the single-layer system in the weak and intermediate correlation regimes 
due to the differentiation between OP and IP 
in a regime of the on-site Hubbard interaction $U$ that includes those estimated for the cuprates. 
The present results are qualitatively consistent with NMR experiments in multi-layer cuprates superconductors.

\section{Formalism} 

\subsection{Three-layer Hubbard model} 
Let us consider a three-layer Hubbard model as a model for the Hg-based three-layer cuprate Hg-1223, 
where the tight-biding parameters are obtained from first-principles calculations. 
The Hamiltonian, 
\begin{equation} 
H = H_{0} +H_{\text{int}} , 
\end{equation} 
is composed of the kinetic part $H_{0}$ and interaction part $H_{\text{int}}$. 
The former is 
\begin{equation} 
H_{0} = -\sum_{ab} \sum_{ij} \sum_{\sigma} t^{ab}_{ij} c^{a\dagger}_{i\sigma} c^{b}_{j\sigma} 
        -\mu \sum_{a i \sigma} n^{a}_{i\sigma} , 
\end{equation} 
where $c^{a \, \dagger}_{i\sigma}$ creates an electron in the $d_{x^{2}-y^{2}}$ orbital at site $i$ on each plane (square lattice) 
with spin $\sigma$ $(=\uparrow, \downarrow)$ in layer $a$ $(=1,2,3)$, 
$t^{ab}_{ij}$ is the single-electron hopping from $(j,b)$ to $(i,a)$, 
$\mu$ denotes the chemical potential, 
and $n^{a}_{i\sigma} = c^{a\dagger}_{i\sigma} c^{a}_{i\sigma}$ is the number operator. 
We here call layers $a=1,3$ the two OPs and layer $a=2$ the IP (see Fig.~\ref{fig:nl}, left panel). 
The intra-layer single-electron hopping is taken into account up to the third-neighbor hopping, 
while the inter-layer single-electron hopping is considered for adjacent layers (i.e., between OP and IP). 
The interaction part is 
\begin{equation} 
H_{\text{int}} = U \sum_{ai} n^{a}_{i\uparrow} n^{a}_{i\downarrow} , 
\end{equation} 
where $U$ is the on-site Coulomb (Hubbard) interaction, 
which is assumed to work within each layer. 

The kinetic part $H_{0}$ can be expressed in a Bloch basis 
$c^{a}_{\bm{k}\sigma} = (1/\sqrt{N}) \sum_i \mathrm{e}^{-i\bm{k} \cdot \bm{R}_i} c^{a}_{i\sigma}$, 
with $N$ being the total number of sites and $\bm{R}_i$ the position of site $i$, as a $3 \times 3$ matrix, 
\begin{equation} 
\begin{split} 
H_{0} 
&=  \sum_{\bm{k}\sigma} \vec{c}^{\, \dagger}_{\bm{k}\sigma} \hat{\xi}_{\bm{k}} \vec{c}_{\bm{k}\sigma}  \\ 
&=  \sum_{\bm{k}\sigma} 
    \left( \begin{array}{ccc} c^{1\dagger}_{\bm{k} \sigma} & c^{2\dagger}_{\bm{k} \sigma} & c^{3\dagger}_{\bm{k} \sigma} \end{array} \right)  \\ 
&\qquad \times 
    \left( \begin{array}{ccc}  \epsilon_{\bm{k}} -\mu  &  t^{\perp}_{\bm{k}}      &  0                       \\ 
                               t^{\perp}_{\bm{k}}      &  \epsilon_{\bm{k}} -\mu  &  t^{\perp}_{\bm{k}}      \\ 
                               0                       &  t^{\perp}_{\bm{k}}      &  \epsilon_{\bm{k}} -\mu  \end{array} \right) 
    \left( \def\arraystretch{1.1} \begin{array}{c}  c^{1}_{\bm{k} \sigma} \\  c^{2}_{\bm{k} \sigma} \\  c^{3}_{\bm{k} \sigma} \end{array} \right) , 
    \label{eq:H0} 
\end{split} 
\end{equation} 
where 
$\vec{c}^{\, \dagger}_{\bm{k}\sigma} = (c^{1\dagger}_{\bm{k} \sigma} \, c^{2\dagger}_{\bm{k} \sigma} \, c^{3\dagger}_{\bm{k} \sigma})$, 
and $\hat{\xi}_{\bm{k}}$ is the energy dispersion matrix defined by the last line. 
The intra-layer energy dispersion $\epsilon_{\bm{k}}$ is 
\begin{equation} 
\begin{split} 
\epsilon_{\bm{k}} 
&= -2t                 \left( \cos  k_{x} +\cos  k_{y} \right)  \\ 
&\quad 
   +4t^{\prime}               \cos k_{x} \cos k_{y} 
   -2t^{\prime \prime} \left( \cos 2k_{x} +\cos 2k_{y} \right) , 
\end{split} 
\end{equation} 
where $t$, $t^{\prime}$, and $t^{\prime \prime}$ represent the intra-layer nearest-, second-, and third-neighbor hopping, respectively, 
while 
\begin{equation} 
t^{\perp}_{\bm{k}} = -t_{\perp} \left( \cos k_{x} -\cos k_{y} \right)^{2} 
\end{equation} 
is the inter-layer single-electron hopping between $d_{x^{2}-y^{2}}$ orbitals via $s$ orbital~\cite{Andersen94,Andersen95,KN13,KN17} 
in the crystal structure shown in Fig.~\ref{fig:nl}. 
These tight-binding parameters can be evaluated from the standard downfolding 
in terms of the maximally-localized Wannier functions derived from the density functional theory (DFT) band calculations,~\cite{KN_thesis} 
where the obtained values for the three-layer cuprate Hg-1223 are 
$(t, t^{\prime}, t^{\prime \prime}, t_{\perp} ) = (0.45, 0.10, 0.08, 0.05)$ eV. 
Other Hg-based multi-layer cuprates Hg-$12(n-1)n$ have similar parameters.~\cite{KN_thesis} 
Although the site potential in the IP evaluated from the DFT band calculations is larger than that for the OP by $\sim 0.1$ eV,~\cite{KN_thesis} 
we have here ignored the difference for simplicity since the effect on the band structure is small. 
One can readily diagonalize the kinetic part $H_{0}$ to have energy eigenvalues $E^{m}_{\bm{k}}$ ($m=1,2,3$), 
where $E^{1,3}_{\bm{k}} = \xi_{\bm{k}} \mp \sqrt{2}t^{\perp}_{\bm{k}}$ and $E^{2}_{\bm{k}} = \xi_{\bm{k}}$, 
with $E^{1}_{\bm{k}} \geq E^{2}_{\bm{k}} \geq E^{3}_{\bm{k}}$ because $t^{\perp}_{\bm{k}} \leq 0$. 
The corresponding field operators are 
$a^{1,3}_{\bm{k}\sigma} = ( c^{1}_{\bm{k}\sigma} \mp \sqrt{2}c^{2}_{\bm{k}\sigma} +c^{3}_{\bm{k}\sigma} )/2$ 
and 
$a^{2}_{\bm{k}\sigma}   = ( c^{1}_{\bm{k}\sigma}                                  -c^{3}_{\bm{k}\sigma} )/\sqrt{2}$. 
The carrier doping is controlled by the chemical potential $\mu$ through Green's function as shown later.

\subsection{TPSC approach for multi-layer systems} 
We next consider the TPSC approach for multi-layer systems.~\cite{KN17} 
The TPSC approach~\cite{Vilk94,Vilk97,Tremblay11,Arita00,Miyahara13,Ogura15,KN17} is developed for weak- and intermediate-coupling regimes 
so as to respect the conservation of spin and charge, the Mermin--Wagner theorem, 
the Pauli principle, the $q$-sum rule for spin and charge susceptibilities, and the $f$-sum rule. 
The spin and charge susceptibilities, along with the double occupancy, 
are determined self-consistently with the TPSC ansatz, 
from which the self-energy and Green's function are evaluated. 
Here we first show how the TPSC approach can be applied to multi-layer Hubbard models. 

We start with Green's function for multi-layer systems defined as 
\begin{equation} 
G^{ab}(k) = -\int^{\beta}_{0} d\tau \, \mathrm{e}^{i\omega_{n}\tau} \bra T_{\tau} c^{a}_{\bm{k}\sigma}(\tau) c^{b\dagger}_{\bm{k}\sigma}(0) \ket , 
\end{equation} 
where $a \, (=1,2,3)$ again denotes the layer index, 
$k=(\bm{k}, i\omega_{n})$ with $\omega_{n} = (2n+1)\pi /\beta$ ($n \in \bm{Z}$) being the Matsubara frequency for fermions, 
$\beta = 1/T$ ($k_{\text{B}}=1$) is the inverse temperature, 
$T_{\tau}$ stands for the imaginary-time ordering, 
and $\bra \cdots \ket$ represents the quantum statistical average. 
Let us consider the spin and charge (orbital) susceptibilities in the multi-layer systems. 
In terms of the spin operators in momentum space, 
$S^{z \, a}_{\bm{q}}= (1/2) \sum_{\bm{k}} ( c^{a \, \dagger}_{\bm{k}\uparrow  } c^{a}_{\bm{k}+\bm{q}\uparrow  } 
                                           -c^{a \, \dagger}_{\bm{k}\downarrow} c^{a}_{\bm{k}+\bm{q}\downarrow} )$, 
$S^{- \, a}_{\bm{q}}= \sum_{\bm{k}} c^{a \, \dagger}_{\bm{k}\downarrow} c^{a}_{\bm{k}+\bm{q}\uparrow}$, 
and 
$S^{+ \, a}_{\bm{q}}= \sum_{\bm{k}} c^{a \, \dagger}_{\bm{k}\uparrow} c^{a}_{\bm{k}+\bm{q}\downarrow}$, 
and the charge operator, 
$n^{a}_{\bm{q}}= \sum_{\bm{k}} ( c^{a \, \dagger}_{\bm{k}\uparrow  } c^{a}_{\bm{k}+\bm{q}\uparrow  } 
                                +c^{a \, \dagger}_{\bm{k}\downarrow} c^{a}_{\bm{k}+\bm{q}\downarrow} )$, 
we define the longitudinal ($zz$) and transverse ($\pm$) spin susceptibilities in a matrix form as 
\begin{equation} 
\begin{split} 
\chi^{ab}_{zz}(q) 
&=  \frac{1}{N} \int^{\beta}_{0} d\tau \, \mathrm{e}^{i\epsilon_{m} \tau} \bra T_{\tau} S^{z \, a}_{\bm{q}} (\tau) S^{z \, b}_{-\bm{q}} (0) \ket ,  \\ 
\chi^{ab}_{\pm}(q) 
&=  \frac{1}{N} \int^{\beta}_{0} d\tau \, \mathrm{e}^{i\epsilon_{m} \tau} \bra T_{\tau} S^{- \, a}_{\bm{q}} (\tau) S^{+ \, b}_{-\bm{q}} (0) \ket , 
\end{split} 
\end{equation} 
and the charge susceptibility as 
\begin{equation} 
\begin{split} 
\chi^{ab}_{\text{C}}(q) 
&=  \frac{1}{N} \int^{\beta}_{0} d\tau \, \mathrm{e}^{i\epsilon_{m} \tau}  
    \frac{1}{2} [ \bra T_{\tau} n^{a}_{\bm{q}} (\tau) n^{b}_{-\bm{q}} (0) \ket -\bra n^{a}_{\bm{q}} \ket \bra n^{b}_{-\bm{q}} \ket ] ,
\end{split} 
\end{equation} 
where $q=(\bm{q}, i\epsilon_{m})$ with $\epsilon_{m} = 2m\pi /\beta$ ($m \in \bm{Z}$) being the Matsubara frequency for bosons. 
In the presence of spin SU(2) symmetry, 
the longitudinal and transverse spin susceptibilities satisfy $2\hat{\chi}_{zz} = \hat{\chi}_{\pm} \equiv \hat{\chi}_{\text{S}}$. 

In the multi-layer TPSC approach, 
the spin and charge susceptibilities are respectively assumed to take the forms,  
\begin{equation} 
\begin{split} 
\hat{\chi}_{\text{S}}(q) = \frac{\hat{\chi}_{0}(q)}{1 -\hat{\chi}_{0}(q) \hat{U}_{\text{S}}} , 
\quad 
\hat{\chi}_{\text{C}}(q) = \frac{\hat{\chi}_{0}(q)}{1 +\hat{\chi}_{0}(q) \hat{U}_{\text{C}}} . 
\label{eq:XSC} 
\end{split} 
\end{equation} 
Hereafter, matrices are displayed with hats, 
and the matrix operations above are defined as 
$[1 -\hat{\chi}_{0} \hat{U}_{\text{S}(\text{C})}]^{-1} \hat{\chi}_{0}$, etc. 
The polarization function $\chi_{0}$ is defined as 
\begin{equation} 
\chi^{ab}_{0}(q) = -\frac{1}{N\beta} \sum_{k} G^{ab}_{0}(q+k) G^{ba}_{0}(k) 
\end{equation} 
from the non-interacting Green's function $\hat{G}_{0}(k) = ( i\omega_{n} -\hat{\xi}_{\bm{k}} )^{-1}$. 
The spin- and charge-channel interactions for the trilayer system, 
\begin{equation} 
\begin{split} 
\hat{U}_{\text{S}} 
&=  \left( \begin{array}{ccc}  U^{\text{OP}}_{\text{S}}  &  0                         &  0  \\
                               0                         &  U^{\text{IP}}_{\text{S}}  &  0  \\
                               0                         &  0                         &  U^{\text{OP}}_{\text{S}}  \end{array} \right) , \quad  \\ 
\hat{U}_{\text{C}} 
&=  \left( \begin{array}{ccc}  U^{\text{OP}}_{\text{C}}  &  0                         &  0  \\
                               0                         &  U^{\text{IP}}_{\text{C}}  &  0  \\
                               0                         &  0                         &  U^{\text{OP}}_{\text{C}}  \end{array} \right) , 
\end{split} 
\end{equation} 
consisting of the OP and IP components, $U^{\text{OP}}_{\text{S}(\text{C})}$ and $U^{\text{IP}}_{\text{S}(\text{C})}$, 
are determined self-consistently along with the double occupancy 
$\bra n^{a}_{\uparrow} n^{a}_{\downarrow} \ket \equiv \bra n^{a}_{i\uparrow} n^{a}_{i\downarrow} \ket$
by the $q$-sum rule for the spin and charge susceptibilities with the TPSC ansatz in multi-layer systems: 
\begin{equation} 
\begin{split} 
\frac{1}{N\beta} \sum_{q} 2\chi^{aa}_{\text{S}}(q) &= \bra n^{a} \ket -2\bra n^{a}_{\uparrow} n^{a}_{\downarrow} \ket ,  \\ 
\frac{1}{N\beta} \sum_{q} 2\chi^{aa}_{\text{C}}(q) &= \bra n^{a} \ket +2\bra n^{a}_{\uparrow} n^{a}_{\downarrow} \ket -\bra n^{a} \ket^{2} , 
\label{eq:qsum} 
\end{split} 
\end{equation} 
and 
\begin{equation} 
U^{\text{OP}}_{\text{S}} = U \frac{\bra n^{1}_{\uparrow} n^{1}_{\downarrow} \ket}{\bra n^{1}_{\uparrow} \ket \bra n^{1}_{\downarrow} \ket} 
                         = U \frac{\bra n^{3}_{\uparrow} n^{3}_{\downarrow} \ket}{\bra n^{3}_{\uparrow} \ket \bra n^{3}_{\downarrow} \ket} ,  \quad 
U^{\text{IP}}_{\text{S}} = U \frac{\bra n^{2}_{\uparrow} n^{2}_{\downarrow} \ket}{\bra n^{2}_{\uparrow} \ket \bra n^{2}_{\downarrow} \ket} . 
\label{eq:ansatz} 
\end{equation} 
Here $\bra n^{a} \ket = \bra n^{a}_{\uparrow} \ket + \bra n^{a}_{\downarrow} \ket$ is the filling in layer $a$, 
where we assume $\bra n^{a}_{\sigma} \ket = \bra n^{a}_{i\sigma} \ket$ (translational symmetry) 
and $\bra n^{a}_{\uparrow} \ket = \bra n^{a}_{\downarrow} \ket$ (paramagnetic state). 

In the TPSC approach, 
the spin susceptibility $\hat{\chi}_{\text{S}}(q)$ is first determined 
along with the spin-channel interaction $\hat{U}_{\text{S}}$ and double occupancy $\bra n^{a}_{\uparrow} n^{a}_{\downarrow} \ket$ 
using the expression for $\hat{\chi}_{\text{S}}(q)$ in Eq.~(\ref{eq:XSC}), 
the $q$-sum rule for $\hat{\chi}_{\text{S}}(q)$ in Eq.~(\ref{eq:qsum}), and the TPSC ansatz Eq.~(\ref{eq:ansatz}), 
where $\bra n^{a}_{\sigma} \ket$ is taken to be the non-interacting one. 
The charge susceptibility $\hat{\chi}_{\text{C}}(q)$ is also determined 
along with the charge-channel interaction $\hat{U}_{\text{C}}$ 
using the expression for $\hat{\chi}_{\text{C}}(q)$ in Eq.~(\ref{eq:XSC}), 
the $q$-sum rule for $\hat{\chi}_{\text{C}}(q)$ in Eq.~(\ref{eq:qsum}), 
and the obtained double occupancy $\bra n^{a}_{\uparrow} n^{a}_{\downarrow} \ket$. 

Once the spin and charge susceptibilities are determined, 
we can obtain the self-energy as 
\begin{equation} 
\begin{split} 
\Sigma^{ab}(k) 
=  \frac{1}{N\beta} \sum_{k^{\prime}} 
    \bigg[ \hat{U} &+\frac{3}{4} \hat{U} \hat{\chi}_{\text{S}}(k-k^{\prime}) \hat{U}_{\text{S}}  \\ 
&+\frac{1}{4} \hat{U} \hat{\chi}_{\text{C}}(k-k^{\prime}) \hat{U}_{\text{C}} \bigg]^{ab} G^{ab}_{0}(k^{\prime}) ,  \label{eq:self-energy} 
\end{split} 
\end{equation} 
where the form of the self-energy (involving products of matrix elements) comes from our assumption 
that the on-site Hubbard interaction only works within each layer, 
and $\hat{U} = \mathrm{diag} \, (U, \, U, \, U)$ denotes the bare on-site Hubbard interaction. 
Then the interacting Green's function in the multi-layer TPSC is given as 
\begin{equation} 
\hat{G}(k) = \Big[ \hat{G}^{-1}_{0}(k) -\hat{\Sigma}(k) \Big]^{-1} . 
\end{equation} 
To evaluate the filling $n^{a}$ in layer $a$, 
and also to determine the chemical potential from the total filling, 
we can use the relation between $n^{a}$ and the Green's function $\hat{G}(k)$,
\begin{equation} 
\begin{split} 
n^{a} 
&= \frac{1}{N\beta} \sum_{k\sigma} \mathrm{e}^{-i\omega_{n} 0^{-}} G^{aa}(k)  \\ 
&= \frac{1}{N} \sum_{\bm{k}\sigma} \left[ \frac{2}{\beta} \sum_{\omega_{n} >0} \mathrm{Re} \, G^{aa}(\bm{k},i\omega_{n}) +\frac{1}{2} \right]. 
\end{split} 
\end{equation} 

Now, superconductivity in multi-layer systems can be studied by the linearized Eliashberg equation for singlet pairings, 
\begin{equation} 
\begin{split} 
\lambda \Delta^{ab}(k) 
&= -\frac{1}{N\beta} \sum_{k^{\prime}} \sum_{a^{\prime} b^{\prime}} V^{ab}_{\textrm{P}}(k-k^{\prime})  \\ 
&\qquad \times 
    G^{aa^{\prime}}(k^{\prime}) \Delta^{a^{\prime} b^{\prime}}(k^{\prime}) G^{bb^{\prime}}(-k^{\prime}).  \label{eq:LEeq} 
\end{split} 
\end{equation} 
Since we deal with multi-layer systems, 
the SC gap function $\hat{\Delta}(k)$ is a matrix 
spanned by the layer indices ($3\times 3$ for a trilayer system). 
There, different components are coupled with each other via the matrix equation, 
so that we have a single eigenvalue $\lambda$. 
The largest eigenvalue can be evaluated numerically by the power-method iteration, 
and the SC transition corresponds to the temperature at which $\lambda$ becomes unity. 
The magnitude of $\lambda$ can also be used as a measure of superconductivity even for $T \gtrsim T_{\text{c}}$.  
The singlet pairing interaction $V_{\textrm{P}}$ can be given in a matrix form as~\cite{Miyahara13,Ogura15} 
\begin{equation} 
\hat{V}_{\textrm{P}}(q) 
=  \hat{U} +\frac{3}{2} \hat{U} \hat{\chi}_{\text{S}}(q) \hat{U}_{\text{S}} 
           -\frac{1}{2} \hat{U} \hat{\chi}_{\text{C}}(q) \hat{U}_{\text{C}} . 
\label{eq:VP} 
\end{equation}

\section{Numerical Results} 
Let us now present the numerical results for SC properties of the three-layer Hubbard model. 
In our calculations, 
the number of discrete mesh of two-dimensional $\bm{k}$-points ($\bm{q}$-points) and Matsubara frequency $\omega_{n}$ ($\epsilon_{m}$) 
are set to be $(k_{x}, k_{y}, \omega_{n}) = (q_{x}, q_{y}, \epsilon_{m}) = (128, 128, 4096)$ throughout, 
with a temperature $T=0.015$ eV ($\sim 150$ K). 

\subsection{Self-doping effect arising from electron correlations} 
\begin{figure}[t]
\begin{center}
\includegraphics[width=8.5cm,clip]{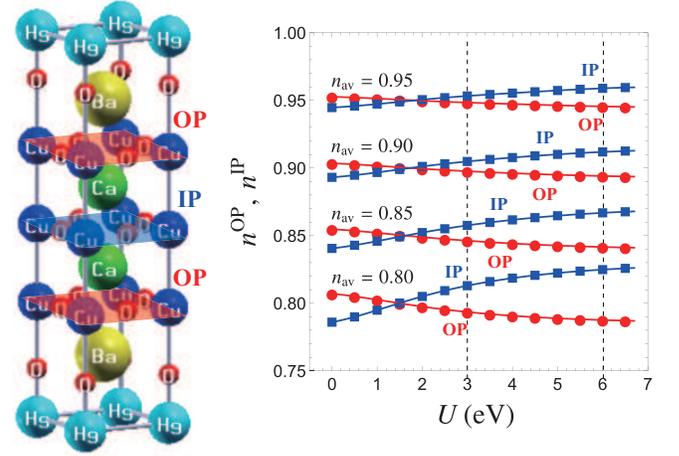} 
\caption{(Color online) 
(Left) Crystal structure of Hg-based three-layer cuprate Hg-1223 with two OPs and one IP in an unit cell. 
(Right) TPSC result for the layer fillings in the OP and IP, $n^{\text{OP}}$ (red circles) and $n^{\text{IP}}$ (blue squares) respectively, 
against the on-site Hubbard interaction $U$ 
for various values of the average filling $n_{\text{av}}= 0.95, 0.90, 0.85, 0.80$. 
Here the vertical dotted lines indicate the values of $U=3.0$ eV and $6.0$ eV which are taken in the following discussions. 
} 
\label{fig:nl}
\end{center} 
\end{figure}

Before we examine the SC properties, 
we need to look at the fillings in the OP and IP, $n^{\text{OP}} \equiv n^{1} = n^{3}$ and $n^{\text{IP}} \equiv n^{2}$, respectively. 
Fig.~\ref{fig:nl} displays $n^{\text{OP}}$ and $n^{\text{IP}}$ against the on-site Hubbard interaction $U$ 
for various values of the average filling $n_{\text{av}} 
\equiv (1/3) \sum^{3}_{a=1} n^{a} = 0.95 - 0.80$. 
For each $n_{\text{av}}$ we can see that 
the filling of OP decreases with $U$, whereas the filling of IP, originally below the OP filling at $U=0$, increases. 
Namely, the two curves cross with each other at a certain $U$, 
causing $n^{\text{IP}}$ exceed $n^{\text{OP}}$ for $U \gtrsim 2.0$ eV. 

This is a self-doping effect arising from the electron correlation:~\cite{KN17} 
In the trilayer system, 
the electrons tend to be redistributed by the differentiation in the electron correlation 
as described by the multi-layer TPSC that determines the self-energy within a one-shot calculation, 
where the electrons (holes) are introduced into the IP (OP). 
The obtained doping behavior is consistent with the NMR experiments,~\cite{Mukuda12,Mukuda08,Mukuda06,Kotegawa01} 
where more hole carriers are shown to be introduced into the OP than IP. 

If we only consider the site potential difference between the OP and IP, 
$\Delta \varepsilon = \varepsilon_{\text{IP}} -\varepsilon_{\text{OP}} > 0$ coming from an effect of the Madelung potential, 
this (with the many-body effect ignored) would transfer the electrons from the IP into OP, 
which is contrary to those observed in the NMR experiments. 
Thus the layer-resolved filling is indeed an electron correlation effect. 

\subsection{SC gap functions in OP and IP}
Now we come to SC properties of the three-layer Hubbard model. 
The present numerical results first confirm that 
the SC gap function $\hat{\Delta}(k)$ that has the maximum eigenvalues of the linearized Eliashberg equation 
is the spin-singlet $d_{x^{2}-y^{2}}$-wave ($\sim \cos k_{x} -\cos k_{y}$) pairing in the parameter range considered here. 
This is natural, 
since the superconductivity is mediated by the AF spin fluctuations,~\cite{Miyahara13,Ogura15} as also elaborated below. 
An essential point for multi-layer systems, however, resides in the fact that 
the SC gap matrix $\hat{\Delta}(k)$ contains off-diagonal (inter-layer) matrix elements 
arising from the inter-layer pairing on top of the diagonal (intra-layer) matrix elements. 
The present calculation shows that 
the amplitude of the inter-layer SC gap functions $\Delta^{ab}(k)$ ($a \neq b$) is much smaller than those for the intra-layer ones, 
$\Delta^{\text{OP}}(k) \equiv \Delta^{11}(k) = \Delta^{33}(k)$ in OPs and $\Delta^{\text{IP}}(k) \equiv \Delta^{22}$ in IP, 
where the ratio of their amplitudes is quantitatively $\Delta^{ab}(k) / \Delta^{\text{OP},\text{IP}}(k) < 10^{-2}$ ($a \neq b$). 
One might then take that the $d_{x^{2}-y^{2}}$-wave superconductivity is isolated within each layer, but this is {\it not} the case: 
We must realize that, in the linearized Eliashberg equation Eq.~(\ref{eq:LEeq}), 
there exists not only the intra-layer pair scattering processes within each layer, 
but also the {\it inter-layer pair scattering} processes via the off-diagonal (inter-layer) elements of the Green function ($G^{ab}$ with $a \neq b$) 
that affects the intra-layer gap functions. 
We may regard the latter process as a kind of microscopic ``inter-layer Josephson coupling" as opposed to macroscopic ones. 

\begin{figure}[t]
\begin{center}
\includegraphics[width=8.5cm,clip]{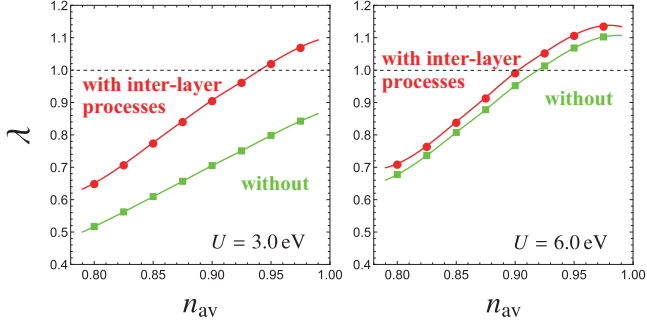} 
\caption{(Color online) 
Eigenvalues of the linearized Eliashberg equation for the three-layer Hubbard model plotted against the average filling $n_{\text{av}}$ 
with (red circles) and without (green squares) the inter-layer processes through the off-diagonal (inter-layer) elements of the Green's function. 
The on-site Hubbard interaction is $U=3.0$ eV (left) and $6.0$ eV (right). 
} 
\label{fig:EV_U}
\end{center} 
\end{figure}

To single out the effect of the inter-layer processes on superconductivity, 
we can look at the effect of artificially switching them off, 
which can be achieved by putting the off-diagonal elements of the Green's function to zero by hand in solving the linearized Eliashberg equation. 
Fig.~\ref{fig:EV_U} displays the eigenvalues of the linearized Eliashberg equation of the three-layer Hubbard model 
with and without the inter-layer processes for various values of the average filling $n_{\text{av}}$. 
The result, here displayed for the on-site Hubbard interaction $U=3.0$ eV and $6.0$ eV, shows that 
the superconductivity is significantly suppressed when the inter-layer scattering processes are switched off. 
Thus the superconductivity is not isolated within each layer, 
but the intra-layer $\Delta^{\text{OP}}(k)$ and $\Delta^{\text{IP}}(k)$ are actually connected with each other 
via the off-diagonal components of the Green's function. 
We can also observe that 
the suppression of superconductivity for the turned-off inter-layer scattering is larger for intermediate $U=3.0$ eV than for a stronger $6.0$ eV.  
We shall clarify the reason below in terms of the weight of the SC gap functions in the OP and IP. 

Now let us look into the layer-resolved SC gap function. 
To quantify the OP and IP components, 
we can define an ``averaged" SC gap function by taking the summation over $k$ (wave number and Matsubara frequency) for each component as 
\begin{equation} 
\bra \Delta^{ab} \ket \equiv \sum_{k} | \Delta^{ab}(k) | , 
\end{equation} 
and for the total average as 
\begin{equation} 
\bra \Delta \ket \equiv \sum_{ab} \bra \Delta^{ab} \ket . 
\end{equation} 
Since the inter-layer components $\bra \Delta^{ab} \ket$ 
($a\neq b$) are much smaller than the intra-layer ones $\bra \Delta^{aa} \ket$ 
according to our numerical results, 
we have only to look at the intra-layer SC gap functions, 
resolved into the OP component 
$\bra \Delta^{\text{OP}} \ket \equiv \bra \Delta^{11} \ket = \bra \Delta^{33} \ket$, 
and the IP component 
$\bra \Delta^{\text{IP}} \ket \equiv \bra \Delta^{22} \ket$. 
Then we can define ``weights" of the OP and IP gap functions as 
$\bra \Delta^{\text{OP}} \ket / \bra \Delta \ket$ and $\bra \Delta^{\text{IP}} \ket / \bra \Delta \ket$, respectively. 

\begin{figure}[t] 
\begin{center} 
\includegraphics[width=7.5cm,clip]{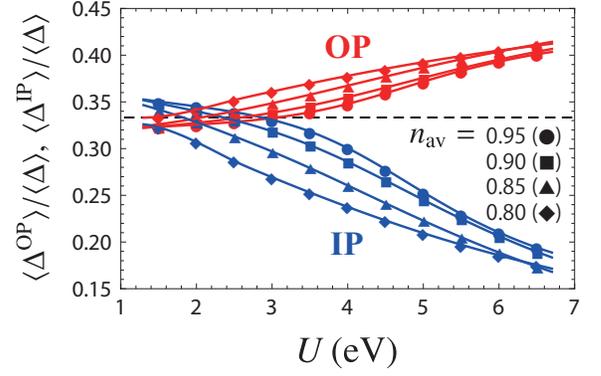} 
\caption{(Color online) 
Weights of the OP and IP gap functions, 
$\bra \Delta^{\text{OP}} \ket / \bra \Delta \ket$ (red symbols) and $\bra \Delta^{\text{IP}} \ket / \bra \Delta \ket$ (blue), 
against the on-site Hubbard interaction $U$ 
for various values of the average filling $n_{\text{av}}= 0.95, 0.90, 0.85, 0.80$. 
The black horizontal line marks $1/3$ (see text).  
} 
\label{fig:SCgap}
\end{center} 
\end{figure}

The numerical result is shown in Fig.~\ref{fig:SCgap} against the on-site Hubbard interaction $U$ 
for various values of the average filling $n_{\text{av}} = 0.95 - 0.80$. 
The black dashed line in the figure marks 
$\bra \Delta^{\text{OP}} \ket / \bra \Delta \ket = \bra \Delta^{\text{IP}} \ket / \bra \Delta \ket = 1/3$, 
which would be the case if the OP and IP had the same averaged amplitudes, 
as would be the case when the inter-layer single-electron hopping is switched off ($t_{\perp} \rightarrow 0$). 

We can see, for each value of $n_{\text{av}}$, that 
the OP component increases with $U$, whereas the IP component decreases. 
This causes $\bra \Delta^{\text{OP}} \ket / \bra \Delta \ket$ 
dominate over $\bra \Delta^{\text{IP}} \ket / \bra \Delta \ket$ for $U \gtrsim 2.0$ eV. 
The crossover $U$ coincides with the crossing of the OP and IP fillings due to the self-doping effect seen in Fig.~\ref{fig:nl}. 
Thus the OP gives a dominant gap function in the $d_{x^{2}-y^{2}}$-wave superconductivity in the three-layer cuprates 
for intermediate or stronger electron correlation. 
The result is qualitatively consistent with the NMR experiments,~\cite{Mukuda12,Mukuda08,Mukuda06,Kotegawa01} 
where the OPs dominate the $d_{x^{2}-y^{2}}$-wave superconductivity for the carrier concentrations up to the overdoped region. 
We can note that, 
while we have taken the linearized Eliashberg equation 
so that we cannot discuss finite amplitudes of the gap functions $\bra \Delta^{\text{OP}} \ket$ and $\bra \Delta^{\text{IP}} \ket$, 
we can still look at their ratio. 
Also, the gap function $\hat{\Delta}(k)$ should not be confused with the eigenvalue $\lambda$ of the linearized Eliashberg equation 
that is related with $T_{\text{c}}$, which will be discussed in Fig.~\ref{fig:EV31L} below. 

The differentiation in the SC gap between the OP and IP found here 
enables us to understand the feature revealed in Fig.~\ref{fig:EV_U}, 
where the suppression of the superconductivity caused by the turned-off inter-layer processes 
is greater for the on-site Hubbard interaction $U=3.0$ eV than for $U=6.0$ eV. 
We can namely interpret this as follows: 
for the weaker $U=3.0$ eV the SC gap functions $\Delta^{\text{OP}}(k)$ and $\Delta^{\text{IP}}(k)$ have really equal weights, 
so that the inter-layer pair scattering processes via the off-diagonal (inter-layer) elements of the Green function 
are relatively important in enhancing the superconductivity, 
while for the stronger $U=6.0$ eV $\Delta^{\text{IP}}(k)$ becomes smaller than $\Delta^{\text{OP}}(k)$ 
so that the contribution of the inter-layer processes to the whole superconductivity becomes less important.

\subsection{Spin susceptibilities and spectral weights in OP and IP} 
\begin{figure}[t] 
\begin{center} 
\includegraphics[width=8.5cm,clip]{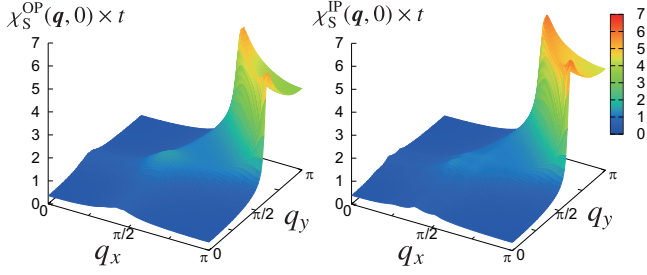} 
\caption{(Color online) 
Spin susceptibilities, $\chi^{\text{OP/IP}}_{\text{S}} (\bm{q}, i\varepsilon_{m}=0)$, 
in the OP (left) and IP (right) for $U=6.0$ eV and $n_{\text{av}}=0.90$. 
The spin susceptibilities are here normalized by the inverse nearest-neighbor hopping $t$ to make them dimensionless. 
} 
\label{fig:XS} 
\end{center} 
\end{figure}

To fathom the multi-layer effects on superconductivity in OP and IP, 
we can go back to the spin susceptibility and spectral weight, 
which are respectively correlated with the pairing interaction $\hat{V}_{\text{P}}(q)$ 
and the quasiparticle DOS which affects the pairing through $\hat{G}(k) \hat{G}(-k)$ in the linearized Eliashberg equation Eq.~(\ref{eq:LEeq}). 
Fig.~\ref{fig:XS} shows the spin susceptibility decomposed into OP and IP, 
$\chi^{\text{OP}}_{\text{S}} \equiv \chi^{11}_{\text{S}} = \chi^{33}_{\text{S}}$ and $\chi^{\text{IP}}_{\text{S}} \equiv \chi^{22}_{\text{S}}$. 
One can see that 
the spin susceptibility around the nesting vector $\bm{Q}=(\pi, \pi)$ in the IP is larger than that in the OP, 
which means that the AF instability in the IP is stronger than in the OP. 
This behavior is seen over the ranges of the on-site Hubbard interaction and the average filling studied here. 
This implies that 
the electron correlation in the IP is stronger than in the OP, 
and is again qualitatively consistent with the NMR experiments,~\cite{Mukuda12,Mukuda08,Mukuda06,Kotegawa01} 
where the AF moments in the IPs are found to be larger than those in the OPs. 

\begin{figure}[t]
\begin{center}
\includegraphics[width=8.0cm,clip]{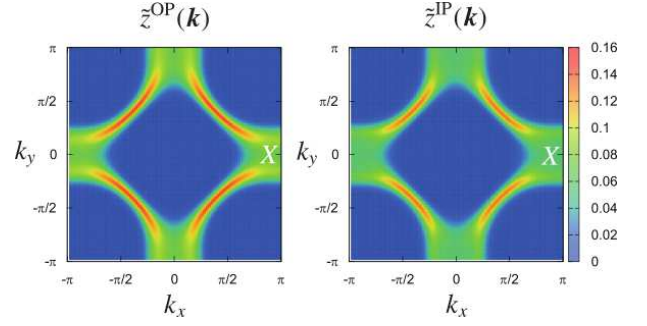} 
\caption{(Color online) 
Averaged spectral weight in the OP (left) and IP (right), 
for $U=6.0$ eV and $n_{\text{av}}=0.90$ as in the previous figure. 
} 
\label{fig:z}
\end{center} 
\end{figure}

\begin{figure}[t]
\begin{center}
\includegraphics[width=8.5cm,clip]{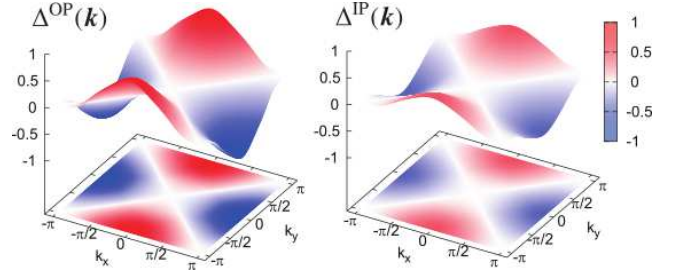} 
\caption{(Color online) 
Gap functions in the OP (left) and IP (right), 
normalized by $\mathrm{max} [ \Delta^{\text{OP}}(\bm{k}, i\pi T )] $ here, 
for $U=6.0$ eV and $n_{\text{av}}=0.90$ as in the previous figures.
} 
\label{fig:OPIP}
\end{center} 
\end{figure}

Since the spin susceptibility, which is related to the $d$-wave pairing interaction, is larger in the IP than OP, 
one might assume that the superconductivity is always favored in the IP than in the OP. 
However, the superconductivity is determined not only by the pairing interaction but also by the quasiparticle DOS. 
So let us look at the averaged spectral weight 
$\tilde{z}$,~\cite{Vilk97,Vilk96,Tremblay11} 
which is obtained from the imaginary-time Green's function 
$G^{ab}(\bm{k}, \tau) = (1/\beta) \sum_{i\omega_{n}} \mathrm{e}^{-i\omega_{n} \tau} G^{ab}(\bm{k},i\omega_{n})$ as  
\begin{equation} 
\tilde{z}^{ab}(\bm{k})  
\equiv -2G^{ab}(\bm{k}, \beta/2) 
=       \int^{\infty}_{-\infty} \frac{d\omega}{2\pi} \frac{A^{ab}(\bm{k},\omega)}{\cosh \left( \beta \omega /2 \right)} , 
\end{equation} 
where $A^{ab}(\bm{k}, \omega) = (-1/\pi) \mathrm{Im} \, G^{\text{R} \, ab}(\bm{k}, \omega)$ 
represents the spectral function evaluated from the retarded Green's function 
$G^{\text{R} \, ab}(\bm{k}, \omega) = G^{ab}(\bm{k}, i\omega_{n}) |_{i\omega_{n} \rightarrow \omega +0^{+}}$. 
Fig.~\ref{fig:z} displays the spectral weight decomposed into OP and IP, 
$\tilde{z}^{\text{OP}} \equiv \tilde{z}^{11} = \tilde{z}^{33}$ and $\tilde{z}^{\text{IP}} \equiv \tilde{z}^{22}$, 
here for $U=6.0$ eV and $n_{\text{av}}=0.90$. 
One can see in Fig.~\ref{fig:z} that 
the averaged spectral weight is reduced especially in the regions around $X$ points in the Brillouin zone called the ``hot spots", 
where the quasiparticle DOS is originally large due to the van Hove singularity and $d_{x^{2}-y^{2}}$-wave superconductivity has large amplitudes. 
If we examine the layer-resolved result, 
the spectral weight in the IP $\tilde{z}^{\text{IP}}$ is seen to be weaker than that in the OP $\tilde{z}^{\text{OP}}$. 
This is always the case for $U \gtrsim 2.0$ eV where the filling in the IP $n^{\text{IP}}$ exceeds that in the OP $n^{\text{OP}}$. 
In other words, 
the averaged spectral weight in the IP is much more suppressed than that in the OP 
due to the strong electron correlation through the self-energy effects, 
since the IP is closer to the half-filing owing to the self-doping effect. 
As a result, 
the gap function in the OP is larger than that in the IP, as mentioned above. 
Fig.~\ref{fig:OPIP} displays the actual gap functions in the OP and IP for $U=6.0$ eV and $n_{\text{av}}=0.90$. 

\subsection{$T_{\text{c}}$ compared between the three- and single-layer Hubbard models} 
\begin{figure}[t]
\begin{center}
\includegraphics[width=7.0cm,clip]{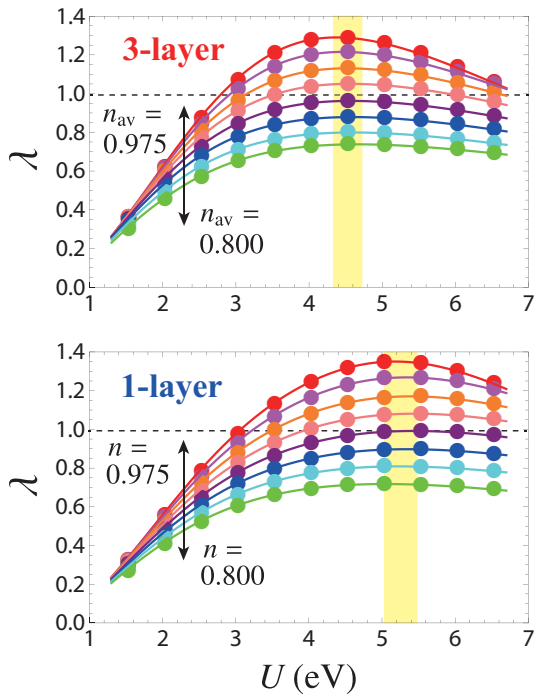} 
\caption{(Color online) 
Eigenvalue $\lambda$ of the linearized Eliashberg equation against the Hubbard interaction $U$ 
for three-layer (top) and single-layer (bottom) Hubbard models  for various values of the average filling 
$n_{(\text{av})} = 0.800$, $0.825$, $0.850$, $0.875$, $0.900$, $0.925$, $0.950$, $0.975$. 
Peak positions are marked with a yellow shading in each panel. 
} 
\label{fig:EV31L}
\end{center} 
\end{figure}

Finally, 
let us compare $T_{\text{c}}$ between the three-layer and Hubbard models 
to identify if and when multi-layer cases can be more favorable for superconductivity. 
Here the single-layer Hubbard model refers to 
the three-layer one without the inter-layer single-electron hopping ($t_{\perp} \rightarrow 0$). 
Fig.~\ref{fig:EV31L} compares the eigenvalue $\lambda$ of the linearized Eliashberg equation for the three-layer and single-layer Hubbard models 
against the on-site Hubbard interaction $U$ for various values of the (average) filling $n_{(\text{av})} = 0.800 - 0.975$. 
We can see that, for each value of $n_{(\text{av})}$, 
the eigenvalues has a dome structure against $U$ for both of the three-layer and single-layer Hubbard models. 
If we have a closer look, however, 
the peak positions (marked with yellow shadings in Fig.~\ref{fig:EV31L}) for the three-layer case 
are shifted to a smaller-$U$ region as compared to those for the single-layer case: 
the peaks of the dome in the three-layer model are located in a range $4.4 \lesssim U \lesssim 4.7$ eV, 
while those for the single-layer model are in a range $5.0 \lesssim U \lesssim 5.5$ eV. 

Let us discuss relevant factors that determine the peak structure for each model. 
The dome structure in Fig.~\ref{fig:EV31L} is determined, 
as we have seen, by the competition between the pairing interaction and quasiparticle DOS. 
The AF spin fluctuations, hence the pairing interaction mediated by them, increases with the on-site Hubbard interaction $U$, 
whereas the quasiparticle DOS decreases with $U$ owing to the increased self-energy, 
so that we can interpret the dome as arising from these two factors having opposite tendencies with $U$. 

\begin{figure}[t]
\begin{center}
\includegraphics[width=8.5cm,clip]{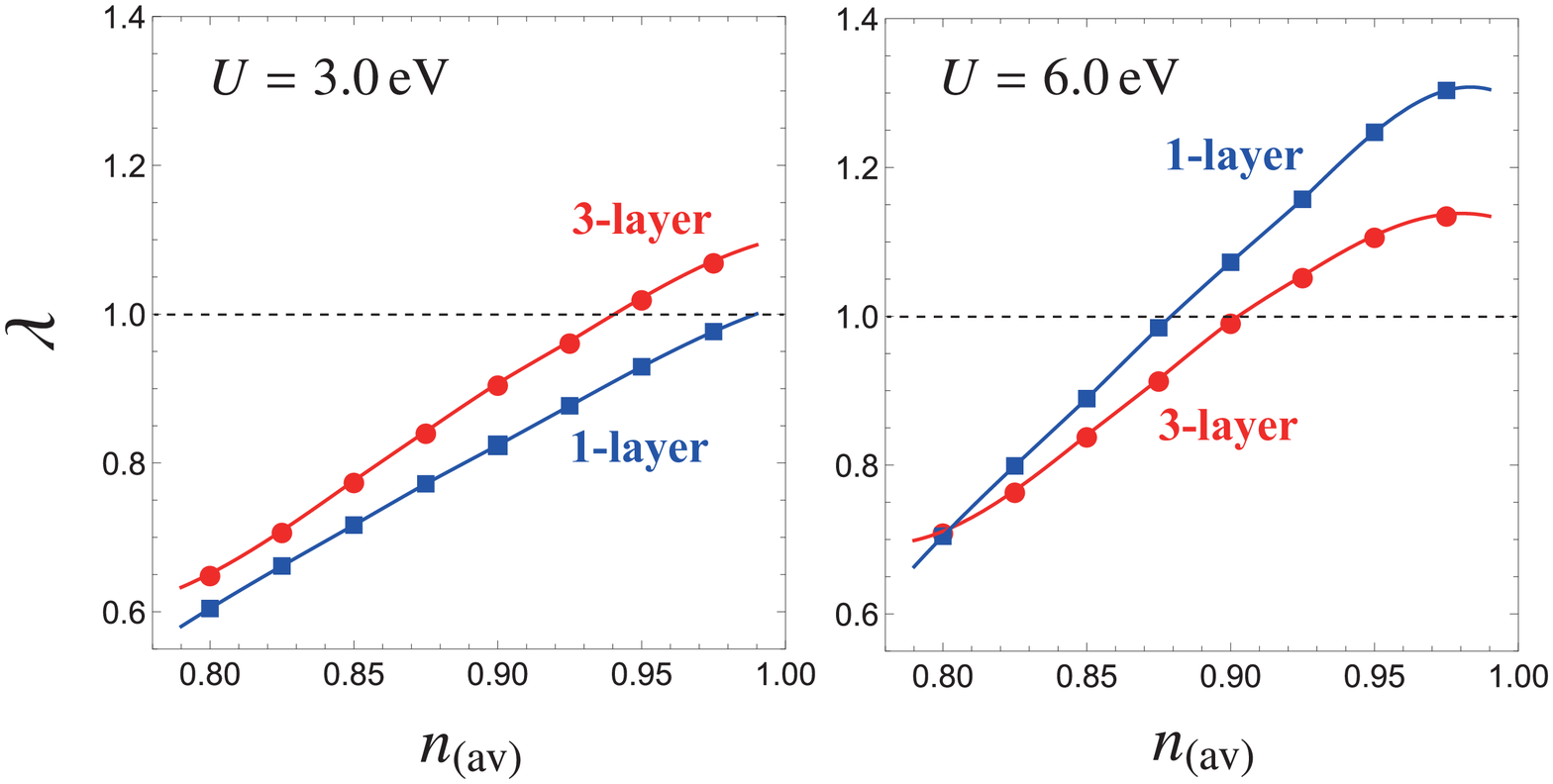} 
\caption{(Color online) 
Eigenvalue $\lambda$ of the linearized Eliashberg equation against the (average) filling 
for the three-layer (red circles) and for the single-layer (blue squares) for the on-site Hubbard interaction $U=3.0$ eV (left) or $U=6.0$ eV (right). 
} 
\label{fig:EVst}
\end{center} 
\end{figure}

One intriguing consequence of the peak positions shifted to a smaller-$U$ region in the three-layer model than in the 
single-layer case is that 
the superconductivity can be {\it enhanced} in the three-layer case than in the single-layer one in a weak $U$ regime. 
In order to quantify this, 
we re-plot in Fig.~\ref{fig:EVst} 
the eigenvalue $\lambda$ of the linearized Eliashberg equation against the average filling $n_{(\text{av})}$ 
to compare between the three- and single-layer models, for the Hubbard interaction $U=3.0$ eV and $U=6.0$ eV. 
One can see that 
the eigenvalues in the three-layer model are indeed significantly enhanced above those in the single-layer model for 
$U \simeq 3.0$ eV. 
Conversely, for a larger $U \simeq 6.0$ eV, 
the $\lambda$ for the single-layer model becomes larger than in the three-layer. 
This is again due to the increased electron correlation (hence the increased self-energy) in the IP compared to OP. 
The $\lambda$s for the three- and single-layer models cross with each other at $U \simeq 4.5$ eV. 

The TPSC approach is a weak- and intermediate-coupling theory, 
and incorporates the self-energy effect arising from the increase of the AF spin fluctuations. 
However, 
the TPSC approach is known to be incapable of describing the Mott transition 
due to the insufficient treatment of local electron correlation,~\cite{Vilk94,Vilk97,Tremblay11} 
so that the fitting lines in Fig.~\ref{fig:EVst} should become invalid toward the half-filling $n_{(\text{av})} \rightarrow 1$. 
In order to extract the true behavior around $n_{(\text{av})} \rightarrow 1$, 
the competition between the SC and Mott insulating phases should be considered. 
As for the eigenvalue $\lambda$  of the linearized Eliashberg equation in Fig.~\ref{fig:EVst}, 
while the numerical results are for a fixed temperature (at $T=0.015$ eV here) with $\lambda$ going below unity in some regions, 
we can still regard $\lambda$ as a measure of $T_{\text{c}}$, 
so that the behavior of $\lambda$ in the three- and single-layer models for $U \simeq 3.0$ eV and $U \simeq 6.0$ eV in Fig.~\ref{fig:EVst} 
should indicate that 
$T_{\text{c}}$ for the three-layer system exceeds that for the single-layer one 
for an intermediate $U \simeq 3.0$ eV, 
while the opposite occurs for a strong $U \simeq 6.0$ eV).

Since the the eigenvalue $\lambda$ ($\sim T_{\text{c}}$) of the linearized Eliashberg equation in Fig.~\ref{fig:EVst} 
almost monotonically increases with approaching the half-filling $n_{(\text{av})} \rightarrow 1$, 
we do not have $T_{\text{c}}$ dome structures observed in experiments, 
which is also due to the insufficient treatment of the electron correlation in the TPSC approach. 
In Fig.~\ref{fig:z} one can see the spectral weight in the regions around the $X$ points tends to vanish, 
i.e., a pseudogap-like behavior. 
Refs.~\cite{Vilk94,Vilk97,Tremblay11} suggested that 
an explicit gap opens in the TPSC approach at lower temperatures.
The origin of this may be regarded as a precursor of AF order 
in that a self-energy effect arising from the divergent behavior of AF spin susceptibility, 
which is considered to be a candidate of pseudogap in electron-doped cuprates,~\cite{Kyung04,Hankevych06} 
while hole-doped cuprates have other candidates including preformed Cooper pairs, competing orders (charge or nematic order), 
and proximity to the Mott insulator.

\section{Summary and Discussions} 
To summarize, 
a three-layer Hubbard model as a model for the cuprate Hg-1223 derived from first-principles calculations 
is studied with the multi-layer TPSC approach for incorporating electron correlations. 
There, the linearized Eliashberg equation for the multi-layer system is solved 
to capture the superconductivity in the OP and IP. 
The present results show that 
the $d_{x^{2}-y^{2}}$-wave superconductivity in the trilayer system 
can be viewed in terms of the different strengths of electron correlation in the OP and IP, 
which is caused by the many-body charge transfer from the OP to IP as a self-doping effect for $U \gtrsim 2.0$ eV: 
the AF spin instability as well as the SC pairing interaction are stronger in the IP than in the OP, 
while for a more strongly correlated regime 
the quasiparticle DOS becomes more suppressed in the IP due to the strong electron correlation than in the OP. 
As a result,  
the OP plays a dominant role in the $d$-wave superconductivity for $U \gtrsim 2.0$ eV, 
which can be grasped from the size of the gap function $\langle \Delta^{\text{OP/IP}} \rangle$ in the OP and IP, 
while the IP dominates the antiferromagnetism, 
as indicated from the strength of the spin susceptibility $\chi^{\text{OP/IP}}_{\text{S}}$. 
The eigenvalue of the linearized Eliashberg equation $\lambda$ ($\sim T_{\text{c}}$), 
which should not be confused with the size of the gap function $\langle \Delta^{\text{OP/IP}} \rangle$, 
becomes peaked around $4.4 \gtrsim U \gtrsim 4.7$ eV, 
where the left of the peak corresponds to the increasing pairing interaction while the right to the blurred spectral function. 
On the other hand, the layer-resolved gap function $\langle \Delta^{\text{OP/IP}} \rangle$ 
serves as a measure of the relative strength of the superconductivity in the OP and IP. 
These electronic and SC properties in the OP and IP, 
i.e., more hole (electron) carriers and stronger superconductivity (antiferromagnetism) in OP (IP), 
are qualitatively consistent with the NMR experimental results. 
We have also shown that 
the three-layer model can have enhanced the $d_{x^{2}-y^{2}}$-wave superconductivity over that in the single-layer model 
for a region of the 
Hubbard interaction $U \lesssim 4.5$ eV, 
while the single-layer one exceeds for $U \gtrsim 4.5$ eV. 

The conclusion that an electron correlation effect leads to the stronger superconductivity (antiferromagnetism) in OP (IP) 
is theoretically obtained here in terms of the self-doping effect that comes from electron correlation 
for the multi-layer Hubbard model with the TPSC approach which is a typical weak- and intermediate-coupling theory. 
This conclusion 
is robust against detailed choice of the values of the parameters within the TPSC approach, although 
the region too close to half-filling with the Mott insulating phase is out of the applicability of the TPSC approach as mentioned above. 
In addition to the TPSC approach, 
we have also investigated the problem with other weak-coupling theories, 
i.e., random phase approximation (RPA) and fluctuation exchange (FLEX) approximation, for the three-layer Hubbard model. 
These methods, however, cannot treat the self-doping effect, 
which is crucial for the three-layer systems as we have revealed in this study. 
For the superconductivity, 
we have employed the effective interaction, Eq.~(\ref{eq:VP}), for the singlet pairing in the Eliashberg equation Eq.~(\ref{eq:LEeq}), 
which is an extension of Ref.~29 and 30 to multi-layer systems. 
The behavior of the effective interaction is qualitatively the same as RPA and FLEX: 
it increases monotonically with increasing on-site Hubbard  $U$ or decreasing temperature $T$. 
However, the linearized Eliashberg equation, which determines SC itself, behaves differently in the TPSC than in the RPA and FLEX. 
This is precisely because the Green's function in the present TPSC approach includes the self-doping effect and disappearance of the spectral weight.  

In the strong-coupling limit, on the other hand, 
a model can be a multi-layer $t$-$J$ model, 
where a possible scenario is that the carriers are redistributed in the OP and IP 
due to a difference in the disrupted exchange interactions between the carrier doping in OP and IP, 
as mentioned in our previous paper:~\cite{KN17} 
holes (electrons) tend to be introduced into the OP (IP) upon doping 
so as to gain the energy from the inter-layer exchange interaction $J^{\perp}$. 
Thus the present result on the self-doping 
seems to encompass both the intermediate and strong-coupling regimes. 
Details on the relation with the $t$-$J$ model, however, will have to be elaborated in the future.

As for the question of which correlation regime the cuprates belong to, 
we can make the following discussion:  
we have indicated in this study that 
$T_{\text{c}}$ is higher for the three-layer cuprate than in the single-layers for $U \lesssim 4.5$ eV. 
Conversely, the single-layer should favor superconductivity for $U \gtrsim 4.5$ eV.  
Then we need an accurate estimate of the on-site Hubbard interaction $U$, including its definition itself in many-body systems. 
A standard numerical method is the constrained random phase approximation (cRPA).~\cite{Springer98,Aryasetiawan04,Aryasetiawan06,Miyake08,Miyake09} 
Recent calculations~\cite{Sasioglu11,Sasioglu13,Jang16} suggest that 
cuprate superconductors lie in a region 
$1.2 \, \text{eV} \lesssim U \lesssim 2.2 \, \text{eV}$ for the Hg-based single-, double-, and triple-layer cuprates 
as far as cRPA is concerned. 
While these estimations suggest that cuprate superconductors sit in a weak-correlation regime, 
the self-doping effect, hence the enhanced $T_{\text{c}}$ for the triple layer case, only takes place for large enough electron correlation. 
Quite recently, it is suggested that 
the cRPA vastly underestimates the size of $U$, 
since the screening arising from the cRPA contribution is canceled by other diagrams.~\cite{Sakakibara17,Honerkamp18}  
This may be relevant to the present study suggesting that 
there exists an intermediate-coupling region ($2.0 \, \text{eV} \lesssim U \lesssim 4.5 \, \text{eV}$) 
where the $d_{x^{2}-y^{2}}$-wave superconductivity is enhanced in the three-layer systems over the single-layer ones. 

Another experimentally-known fact is 
that hydrostatic pressure can increase $T_{\text{c}}$ in cuprate superconductors, 
typically in multi-layer cuprates.~\cite{Schilling93,Takeshita13,Yamamoto15} 
An obvious effect of the pressure is lattice compression within each layer, 
which implies an increased band width $W$, hence a decreased $U/W$, 
and the electron correlations should be decreased by pressure as far as this factor is concerned. 
If we turn to the $T_{\text{c}}$-dome structure against the on-site Hubbard interaction $U$, 
the peak region sits around $U=4.0$--$5.0$ eV in the present study, 
as well as in many theoretical literatures.~\cite{Ogata08,Yokoyama13,Yanagisawa16} 
This implies that 
$T_{\text{c}}$ should rather decrease with pressure 
if we start from the left of the peak (e.g., $U \simeq 2$ eV) as estimated by the cRPA at ambient pressure. 
This contradiction may suggest that here, too, the underestimated $U$ in the cRPA has to be reconsidered. 
Also, microscopic pressure effects other than the lattice reduction have to be considered as shown by Sakakibara {\it et al} 
in a model 
that incorporates $d_{x^{2}-y^{2}}$ main orbital along with the $d_{z^{2}}$ orbital.~\cite{Sakakibara10,Sakakibara12,Sakakibara14} 
They point out that, 
while the change in the band width $W$ is indeed a large effect, we also have 
a pressure effect on the Fermi surface nesting through a change in the second and further neighbor transfers, 
and a pressure effect on the level offset (band spacing) between the copper $d_{x^{2}-y^{2}}$ and $d_{z^{2}}$ orbitals. 
So we may have to consider these factors, on top of, or in relation to, the self-doping effect arising from electron correlations discussed here. 

Experimentally, uniaxial pressure effects~\cite{Hardy10} may give further insights. 
Recently, $T_{\text{c}}$ in multi-layer cuprates is reported 
to be increased not only by $a$-axial (in-plane) pressure but also by $c$-axial (out-of-plane) one.~\cite{Mito12,Mito14,Mito16,Mito17} 
The result would suggest that, as the in-plane (out-of-plane) pressure induces the $a$-axis compression (expansion), 
the effective strength $U/W$, hence $T_{\text{c}}$, 
may possibly change oppositely for the in-plane and out-of-plane pressures. 
A first-principles band calculation~\cite{Mito17} for Hg-based three-layer cuprate suggests that 
the pressure effects may cause a self-doping effect from the HgO block to the CuO$_{2}$. 

In another avenue, 
the SC enhancement due to the differentiation of OP and IP 
reminds us of the artificial superlattices considered in the previous studies,~\cite{Okamoto08,Chen10} 
where a multi-layer Hubbard or $t$-$J$ models composed of overdoped and underdoped layers. 
There, with layer fillings fixed by hand, 
electron correlation effects are investigated with strong-coupling theories 
such as the cellular dynamical mean-field theory, dynamical cluster approximation, 
slave-boson mean-field theory, and Gutzwiller-projected mean-field approximation. 
If we regard the overdoped (underdoped) layer corresponding to the OP (IP) in the present study, 
the SC enhancement found in their previous studies may have some relevance to the present result. 
However, we have to again recall that, 
while the layer filling is fixed in the above studies, 
the fillings of IP and OP are self-consistently determined by the electron correlation as the present paper reveals. 

The high-$T_{\text{c}}$ cuprates are known to accommodate, besides superconductivity and antiferromagnetism, 
various quantum phases such as density-wave and pseudogap phases, 
and extension of the present study to those will be another interesting future work.

\section*{Acknowledgments}
We wish to thank N. Takeshita (AIST) and M. Mito (Kyushu Institute of Technology) for illuminating comments on experimental pressure effects.  
KN would also like to acknowledge Advancesoft Co. for a support from the project ``Study of a simulation program for the correlated electron systems", 
and JSPS KAKENHI Grant No. JP26400357. 
HA acknowledges Shin-ichi Uchida for valuable discussions, 
and JSPS KAKENHI Grant No. JP26247057 and ImPACT Program of Councilfor Science, Technology and Innovation, Cabinet Office, 
Government of Japan (Grant No. 2015-PM12-05-01) for support.  
HA also wishes to thank the Department of Physics, ETH Z\"{u}rich, Switzerland, for hospitality when the manuscript was started. 
Numerical calculations were done at the ISSP Supercomputer Center of the University of Tokyo.


%

\end{document}